\documentclass[conference]{IEEEtran}
\usepackage{layout}
\IEEEoverridecommandlockouts 

\ifCLASSINFOpdf

\else
\fi
\usepackage{caption}
\usepackage{subcaption}
\usepackage{mathtools}
\usepackage{amsmath}
\usepackage{amssymb}
\usepackage{cite}
\usepackage{graphicx}
\usepackage{siunitx}
\usepackage{color}
\usepackage{xcolor}

\begin{document}
\title{Designing A Dimmable OPPM-Based VLC System Under Channel Constraints}
\author{\IEEEauthorblockN{Ata Chizari$^\dag $, Mohammad Vahid Jamali$^\dag $, Jawad A. Salehi$^\dag $,~\IEEEmembership{Fellow,~IEEE}, and Akbar Dargahi$^\ddag $}
\IEEEauthorblockA{$^\dag $ Optical Networks Research Laboratoy (ONRL), Sharif University of Technology (SUT), Tehran, IRAN\\
$^\ddag $Department of Electrical Engineering, Shahid Beheshti University (SBU), Tehran, IRAN \\
Email: \{chizari.ata, mohammad.v.jamali\}@gmail.com, jasalehi@sharif.edu, a-dargahi@sbu.ac.ir\\
}}

\maketitle
\begin{abstract}
In this paper, we design a dimming compatible visible light communication (VLC) system in a standard office room according to illumination standards under channel constraints. We use overlapping pulse position modulation (OPPM) to support dimming control by changing the code weights. The system parameters such as a valid interval for dimming together with an upper bound for bit rate according to the channel delay spread are investigated.
Moreover, considering the dispersive VLC channel and using Monte Carlo (MC) simulations, a method is proposed to determine the minimum code length in different dimming levels in order to achieve a valid bit error rate (BER).
  Finally, trellis coded modulation (TCM) is suggested to be applied to OPPM in order to take advantage of consequent coding gain which could be up to $3$ dB.

\end{abstract}

\IEEEpeerreviewmaketitle
\vspace{-0.2in}
\section{Introduction}
Visible light communication (VLC) is an attractive field of optical wireless communications (OWC) that exploits visible light emitting diodes (VLEDs) as optical sources. This technology which has emerged in recent years, uses the visible band of the electromagnetic spectrum. As a result, it offers an immense unregulated bandwidth for communication in the presence of illumination. Furthermore, as light does not penetrate through walls, secure communication along with enabling frequency reuse between two rooms would be possible. Moreover, since the active area of photo-detectors (PDs) used in VLC is thousands of times greater than the wavelength of light, multipath fading does not occur, however, multipath distortion still can exist in such systems \cite{kahn1997wireless}.

The demand for wireless bandwidth capacity as predicted by Cisco (Cisco VNI. Feb 2015) shows a 10 times growth in mobile traffic for the next five years. On the other hand, during the same years, the mobile carriers are predicted to be accelerated by $9\%$ \cite{chi2015visible}. In addition, most of mobile traffic takes place in indoor environments and at fixed locations. This makes VLC a highly probable technology for wireless access networks of the future home and office.

As mentioned earlier, VLC provides illumination in parallel to communication. Additionally, flicker mitigation and intensity control, also known as dimming control, are two noticeable requirements of such systems according to IEEE 802.15.7 task group \cite{rajagopal802tg7}. Thus, utilizing robust modulation schemes that supply high data rate along with dimming control are imperative. Some dimming adaptable procedures for flicker-free high data rate VLC systems based on IEEE 802.15.17 standard are studied in \cite{ata}.

Among different kinds of modulation schemes, pulse position modulation (PPM) and its families are appropriate for intensity modulation with direct detection (IM/DD) communication systems such as VLC, since the chips within a code word can directly modulate a driver at the transmitter side. In recent years, many research interests have been attracted to overlapping PPM (OPPM) and its applications due to its high spectral efficiency and low bandwidth requirement. In \cite{bai2010joint} a method for supporting dimming by changing the amplitude of OPPM symbol pulses is proposed. Nevertheless, this may result in undesired chromaticity shift of the emitted light due to the characteristics of LEDs \cite{dyble2005impact}. Researchers in \cite{lee2011modulations} proposed multiple PPM (MPPM) to support dimming while transmitting data stream. They explored that MPPM outperforms variable PPM (VPPM) and variable on-off keying (VOOK), in terms of spectral efficiency and power requirement. Moreover, a solution to address the dimming control along with data transmission is suggested in \cite{gancarz2015overlapping} by changing OPPM code word weight.

{This research is inspired by the need to design a VLC system with more spectral efficiency and less power requirement}. In this paper, regarding dimming support by changing code weights of OPPM symbols \cite{gancarz2015overlapping}, we consider a practical scenario within a standard room, and we design a system according to constraining parameters. To do so, we first determine an interval for dimming percentage in which the illumination standards are taken into account. Then, we calculate the maximum data rate for inter-symbol interference- (ISI-) free transmission by simulating the dispersive channel response. Next, we propose a method to determine the maximum usable code length corresponding to a maximum bit error rate (BER) in the presence of different brightness percentages. Owing the fact that the modification in modulation and using coding schemes are essential to performance enhancement of VLC systems \cite{lee2015modulation}, we suggest to apply trellis coded modulation (TCM) in order to take advantage of the corresponding coding gain which results in an improvement in the power requirement of OPPM. Also, the validity of such application is depicted by simulation results.

\section{System and Channel Model}
\subsection{System model}
VLC systems use IM/DD transmission approach since it is an efficient technique in short range indoor wireless optical applications. In this work, we assume the non-directed line-of-sight (LOS) VLC link in which transmitters and receivers have a wide range of transmission and field of view (FOV), respectively (see Fig. 1). Considering the baseband channel model for optical system, the photo-current generated by the receiver PD is given by;
\begin{equation}\label{eq1}
Y(t)=RX(t)*h(t)+n(t), 
\end{equation}
 where $R$ denotes the PD responsivity, $X(t)$ is the transmitter optical power, $*$ represents the convolution operator, and $n(t)$ is the receiver noise. This noise can be modeled as a signal dependent additive white Gaussian noise (AWGN) with double sided power spectral density of $N_0$ as \cite{komine2004fundamental};
 \begin{equation}\label{eq2}
N_0= \sigma _{shot}^{2} +\sigma _{th}^{2}+(RP_r^{(\rm ISI)})^2,
 \end{equation}
in which $\sigma _{shot}^{2}$, $\sigma _{th}^{2}$, and $P_r^{(\rm ISI)}$ are shot noise variance, thermal noise variance, and the received power due to multipath reflection, respectively\footnote{{Note that based on comprehensive study of \cite{komine2004fundamental}, the term $\sigma _{shot}^{2}$ itself takes both incoming optical signal (desired and undesired interfering signals) and ambient background light into account.}} \cite{komine2004fundamental}.
Moreover, the fact that $X(t)$ represents the instantaneous optical power, necessitates $X(t)>0$ as well as $\mathop{\lim }\limits_{T\to \infty }\! \frac{1}{2T}\!\! \int _{-T}^{T}X(t)dt\!<\!P$, where $P$ is the average optical power constraint of the transmitter LED \cite{kahn1997wireless}.  

\begin{figure}\label{1}
       \centering
       \includegraphics[width=3.4in]{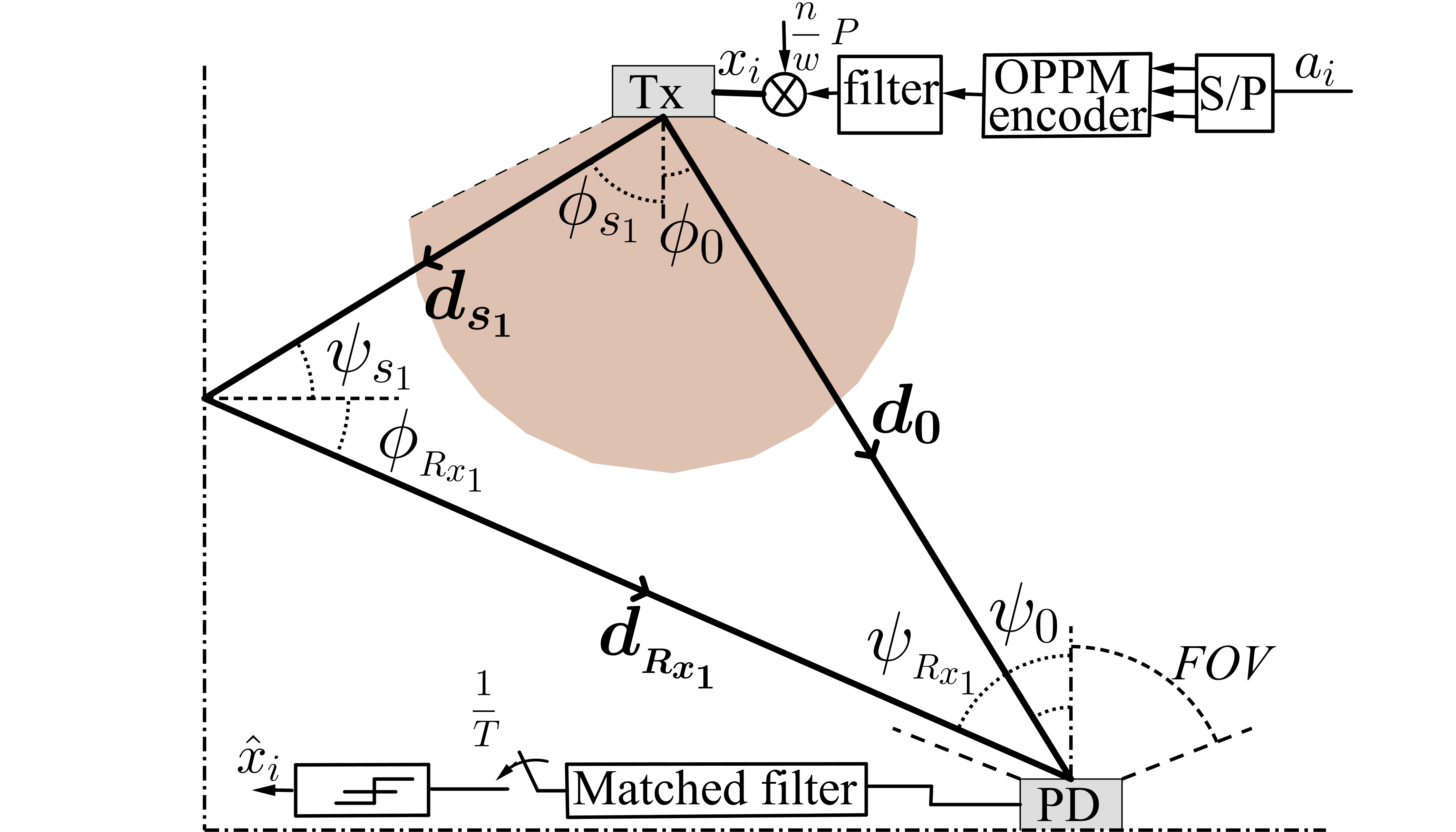}
       \caption {\small System and channel model where Tx, and PD stand for driver and LED array, and photo-detector circuit, respectively.}
                   \vspace{-0.1in}
  \end{figure}
\subsection{Dimming Support in OPPM}  
Consider an $L$-ary modulation scheme in the presence of AWGN and the maximum likelihood (ML) sequence detection, as well. The $L$ non-negative symbols $\{x_1(t), x_2(t),...,x_L(t)\}$ are sent with the bit rate $R_b$ every $T={\log_2^L}/{R_b}$ seconds. The average signal power is\cite{park1995modulation};
\begin{equation} \label{eq3} 
P=\frac{1}{L}\sum _{i=1}^{L}\left\langle x_{i} (t)\right\rangle,   
\end{equation} 
 where $\left\langle .\right\rangle =\mathop{\lim }\limits_{T\to \infty } \frac{1}{2T} \int _{-T}^{T}x(t)dt $. Assuming high signal to noise ratio (SNR), the BER is dominated by two nearest signals and is given by $ Q({d_{min}}/{\sqrt{4N_0}}) $, where $d_{min}^2=\mathop{\min }\limits_{i\ne j} \int (x_{i} (t)-x_{j} (t)) ^{2} dt$ is the minimum Euclidean distance between any pairs of valid symbols.

Looking at Fig. 1, the electrical bit sequence, $a_i$, is turned to parallel to be modulated by OPPM encoder. Then, after passing through filter and being normalized, the resulted modulated symbols, $x_i$, will be sent to block Tx, containing driver circuit and LED arrays. In the case of OPPM, the symbol interval $T$ is divided to $n$ chips and the temporal signal equation would be;
\noindent 
\begin{equation} \label{eq4} 
x(t)=\frac{P}{w} \sqrt{nT} \, \sum _{k=0}^{n-1}c_{k} \sqrt{\frac{n}{T} } p(t-k\frac{T}{n} ),  
\end{equation} 
where ${c}_{k}$s, for $k=0,1,...,n-1$, are binary $n$-tuples of weight $w$, and $p(t)$ is a rectangular pulse with unit amplitude in the interval $[0,T/n]$. Note that $x(t)$ is always positive and fulfills the power constraint mentioned in \eqref{eq3}. The $w$ ones are restricted to be successive whereby the bandwidth requirement in this modulation scheme becomes smaller compared to that of $L$-PPM or MPPM. This advantage is offset by a reduction in the number of OPPM alphabet size. The number of symbols in OPPM is $L=n-w+1$ and the bandwidth requirement is given by;
\begin{equation} \label{eq4.5}
B^{(\rm OPPM)}=\frac{n/w}{{\log_{2}^L}/{R_{b} }}.
\end{equation} 
 Hence, the maximum achievable bit rate and the spectral efficiency can be respectively calculated as follows \cite{park1995modulation};
\begin{equation} \label{eq5} 
\begin{array}{l}{R_{b}<B^{(\rm OPPM)}\Delta \log_{2}^L}, \\ \\{SE=\frac{1}{\Delta}\log_{2}^L}, \end{array} 
\end{equation} 
where $\Delta=w/n$ denotes the duty cycle. Now, in contemplation of controlling the intensity of incident light, one can change the weight of symbols, $w$, while remaining the number of chips, $n$, unchanged \cite{gancarz2015overlapping}. This technique is shown in Fig. 2 in which four dimming levels are illustrated. Obviously, in the cases of $\Delta=1$ and $0$, i.e., full brightness and full darkness, respectively, no data could be transmitted. We also refer to $100\sqrt{\Delta} \%$  as the perceived brightness percentage in that it is more compatible with the nonlinear response of the human eye to the linear changes in the intensity of visible light \cite{ata}. 
\begin{figure}\label{2}
       \centering
       \includegraphics[width=3.4in]{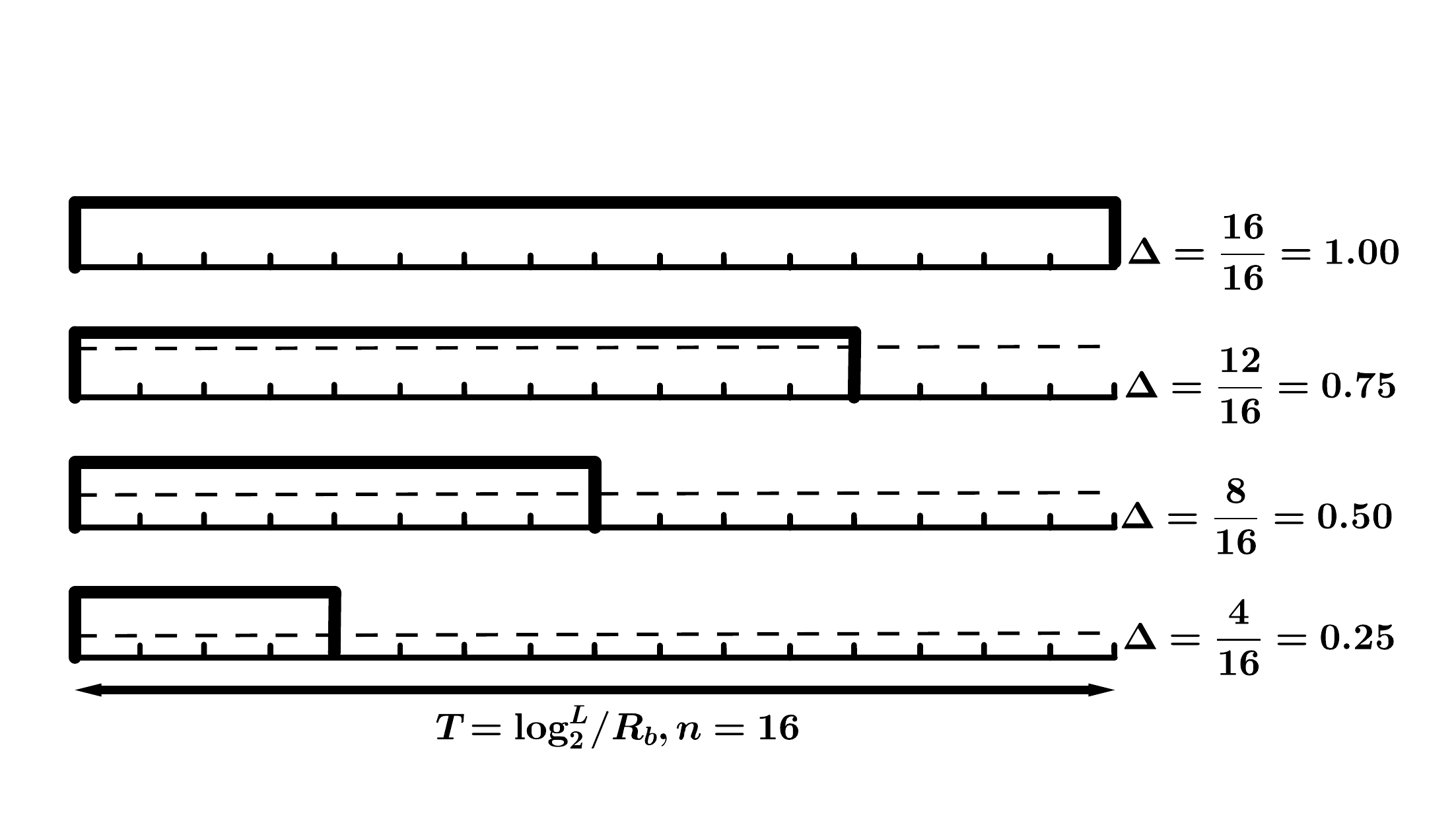}
       \caption {\small Dimming control approach in OPPM, by changing the code weight. Dashed lines represent the transmitted optical power $P$.}
                   \vspace{-0.1in}
  \end{figure}
\subsection{Channel Model}
The photo-current generated by PD is proportional to the received optical power; therefore, the SNR in a VLC link is proportional to the square of received average optical power. This issue makes VLC system vulnerable to the link length, i.e.,  by increasing the distance between the transmitter and the receiver, the SNR will significantly decrease due to the path loss. The SNR is given by \cite{kahn1997wireless};
\begin{equation} \label{eq6} 
SNR=\frac{(RP^{{(\rm LOS)}}_{r} )^{2} }{N_{0} B^{(\rm OPPM)}} = \frac{(RH^{(\rm LOS)}(0)P_{t} )^{2}} {N_{0} B^{(\rm OPPM)}},  
\end{equation}
where $t$ and $r$ stand for transmitted and received terms, respectively. $H^{(\rm LOS)}(0)$ is the portion of channel DC gain arising from the LOS link. The total channel DC gain resulted from both LOS and non-LOS (NLOS) links is defined as $H(0)=\int _{-\infty }^{+\infty }h(t)dt $. Furthermore, the transfer function $H(f)$ would be $H^{(\rm LOS)}(f) +H_{dif} (f)$, in which the contribution due to the LOS link depends on the distance between the transmitter and the receiver, $d_0$, as illustrated in Fig. 1; while the second term relies on modulation frequency besides the room dimensions. The impulse response in the time domain is defined as $H^{(\rm LOS)}(0)\delta (t-{d_{0} }/{c} )$, where $\delta(.)$ represents Dirac delta function and $c$ is the light velocity. Additionally, the channel DC gain is \cite{ghassemlooy2012optical};
\begin{align} \label{eq7}
H^{(\rm LOS)}(0)=\left\{\begin{matrix}
A_r\frac{m+1}{2\pi d_0^2}{\cos}^m(\phi_0)T_s(\psi_0)\\
\times g(\psi_0)\cos(\psi_0), & 0\leq\psi_0\leq FOV \\ 
0, & {\rm Otherwise}
\end{matrix}\right.
\end{align}
where $A_r$ denotes the PD active area, $m$  is Lambert's mode number of a radiation lobe, $-1/{\log_2({\cos(\phi_{1/2})})}$ in which $\phi_{1/2}$ represents the semi-angle at half power of an LED, $T_s(.)$ and $g(.)$ represent the optical bandpass filter transmission and non-imaging concentrator gain of the receiver, and $\phi_0$ and $\psi_0$ are the angle of irradiance and incidence, respectively, as shown in Fig. 1. Thus, the received optical power from LOS path is $P_r^{(\rm LOS)} =H^{(\rm LOS)}(0)P_{t}$.

For the sake of simplicity, we consider only the first reflection which is more dominant than the higher order reflections. So, the NLOS channel response is given by \cite{ghassemlooy2012optical};
\begin{align} \label{eq8} 
h^{l}(t,\!S,\!R_{x})\!=&\!\sum_{j=1}^{\Re}\!\frac{(m\!+\!1)\rho _{j} A_{r}\Delta A}{2\pi d_{s_{j}}^{2}d_{R_{x_{j}}}^{2}}\!\cos^{m}(\phi_{s_{j}}\!)\!\cos(\psi_{s_{j}}\!)T_{s}(\psi_{R_{x_{j}}}\!)\nonumber \\
&\times g(\psi_{R_{x_{j} } }\!)\cos (\phi _{R_{x_{j} }}\!)\cos(\psi _{R_{x_{j} }}\!)\delta (t\!-\!\frac{d_{s_{j} }\!\!+\!d_{R_{x_{j} } } }{c} )\nonumber\\
=& H_{\rm NLOS} (0)\delta (t-\frac{d_{s_{j} } +d_{R_{x_{j} } } }{c} ),
\end{align}
 where $S$ represents the reflectors' surface (consists of four walls in the hypothetical room), $\rho_j$ is reflection index of the $j${th} reflector, $\Delta A$ is the element area of reflectors, and $\Re$ denotes the number of reflectors, i.e., walls. For a diffused link scenario, two components are involved. First, elements of surfaces receive light from transmitter LED. Second, these elements re-emit a portion of received light to PD, as depicted in Fig. 1.  

\section{System Design and Simulation results}
In this section, we design a system according to what was introduced in Section II. To do so, we first investigate a valid interval for dimming by which the illumination standards within a room are taken into account. Then, we obtain the maximum bit rate at which the ISI could be neglected and we compare it with the maximum achievable bit rate offered by OPPM modulation, at a predetermined modulation bandwidth, with respect to different dimming levels. Afterward, we propose a well-known coding scheme, namely TCM, to be applied to OPPM in order to take advantage of the offered coding gain. Finally, from the BER point of view, the minimum required code length to reach a maximum probability of error regarding channel constrains are investigated using MC simulations.

With respect to designing a dimming compatible indoor VLC system assuming practical situation of an office room, we contemplate a $ 5\times 5\times 3$ ${\rm m^3}$ room having four fixtures of LEDs as well as a table of height $0.85$ m under one of the fixtures with a PD on its center. This scenario is shown in Fig. 3. Furthermore, the specification of transmitters and receivers utilized in simulations is summarized in Table 1 where some of LED and PD parameters are extracted from \cite{grubor2007high}.
\begin{figure}\label{3}
       \centering
       \includegraphics[width=3.4in]{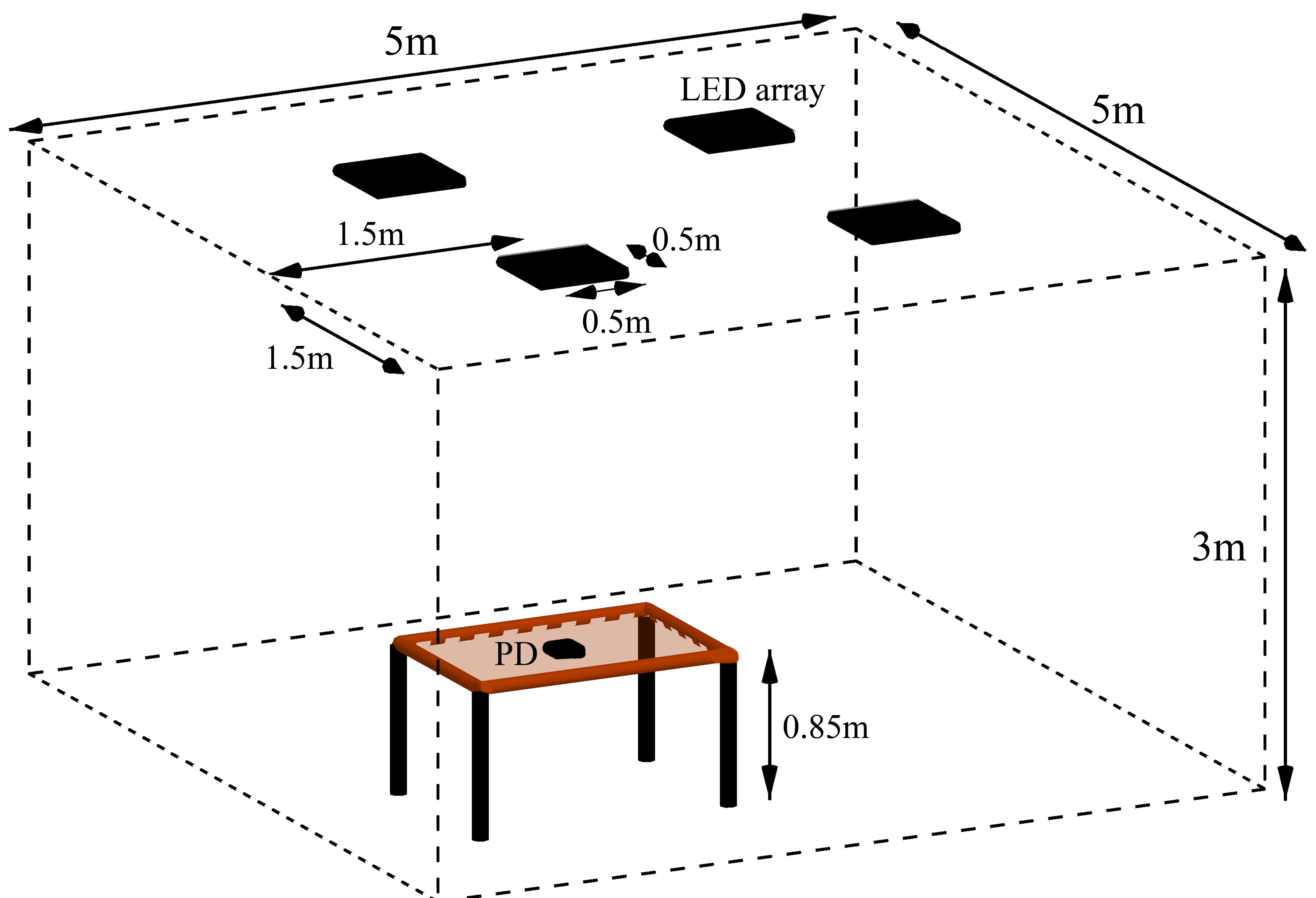}
       \caption {\small The hypothetical office room containing four Landmarks and a PD.}
                   \vspace{0in}
  \end{figure}

\begin{table}[t]
\centering
\caption{\small System parameters.}
\label{tbl:system parameters}
\begin{tabular}{|l|l|}
\hline
\multicolumn{2}{|c|}{{\bf Transmitter side}}                \\ \hline \hline
Number of LED fixtures          & $4$                         \\ \hline
LED fixture area                & $0.5\times 0.5$ ${\rm m^{2}}$     \\ \hline
Number of LEDs within a fixture & $324$                       \\ \hline
LED spacing in a fixture        & $2.8$ cm                  \\ \hline
LED power                       & $63$ mW                   \\ \hline
Semi-angle at half power        & $70 ^{\circ}$                \\ \hline
Center luminous intensity       & $9.5$ cd                  \\ \hline
\hline
\multicolumn{2}{|c|}{{\bf Receiver side (Photo-detector)}}  \\ \hline \hline
Position                        & ${(1,1,0.85)}$ m        \\ \hline
Area                            & $1$ $\rm {{cm}^{2}}$               \\ \hline
FOV                             & $60 ^{\circ}$                \\ \hline
Ambient light current          & $27$ mA                    \\ \hline
Responsivity                    & $0.28$ A/W                 \\ \hline
\end{tabular}
\end{table}
\subsection{Illumination Standards}
The main goal of VLC systems is to provide illumination as well as communication. Therefore, in an office environment the illumination generated by LEDs must satisfy illumination standards. The horizontal illuminance resulted by OPPM modulation with dimming support on a surface is given by;
\begin{equation} \label{eq9} 
E_{h} =\frac{w}{n}\frac{I_{0} \cos ^{m} \left(\phi _{0} \right)\cos \left(\psi _{0} \right)}{d_{0}^{2} },  
\end{equation} 
where $I_0$ represents the center luminous intensity of LEDs. According to \cite{grubor2007high}, the standard illuminance level within an office room is between $200$ and $800$ lux. The variation of horizontal illuminance in terms of different dimming levels in the hypothetical room is depicted in Fig. 4. Note that this figure is obtained at the position of PD shown in Fig. 3. Hence, concerning the illumination budget, a range of $44$ to $90$ percent of perceived brightness would be achievable. Moreover, the distribution of illuminance within the room at $80\%$ of perceived brightness is illustrated in Fig. 5. It is worth to be mentioned that the acquired interval for dimming, fulfills the standard around the table on which the PD is installed.

\begin{figure}\label{4}
       \centering
       \includegraphics[width=3.4in]{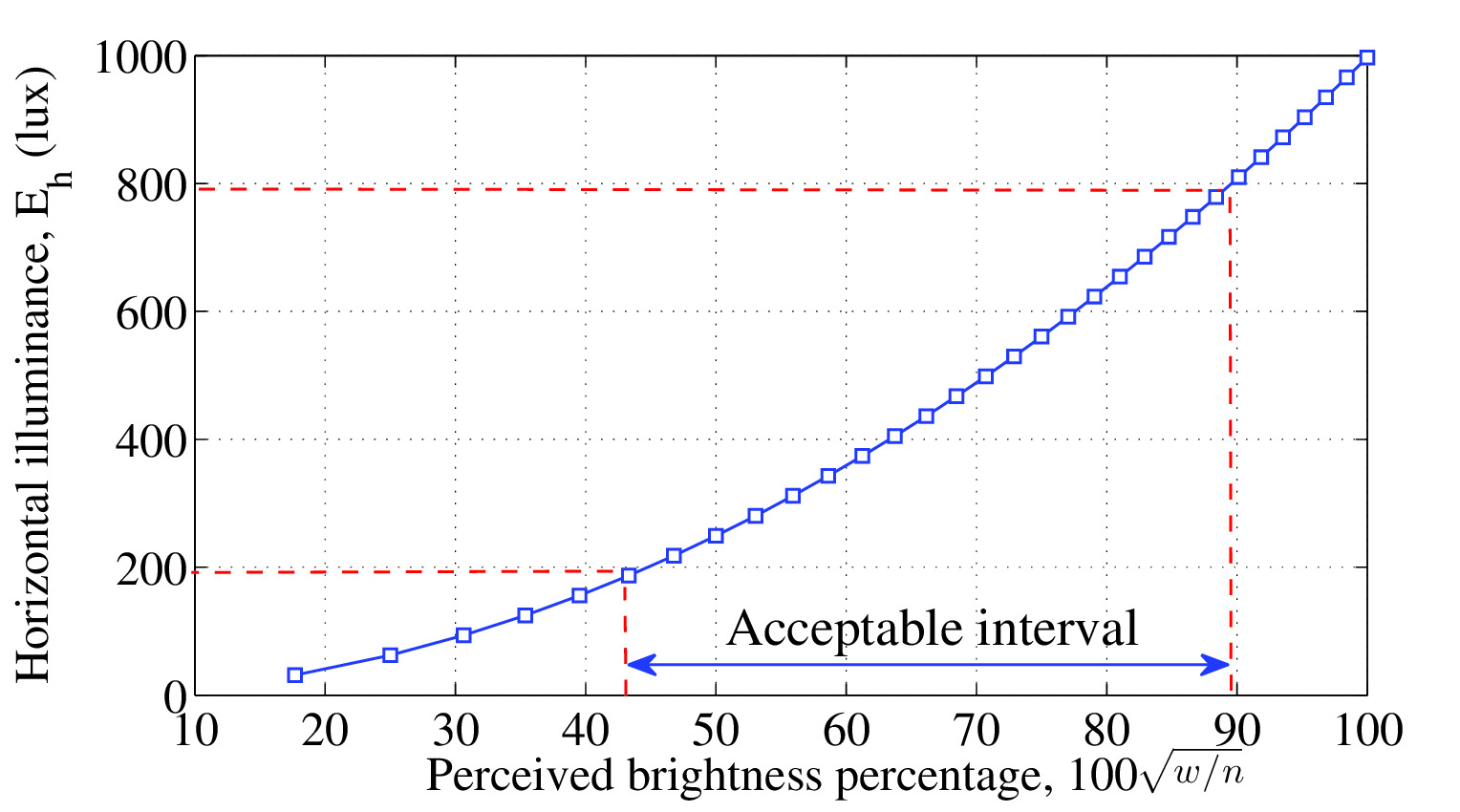}
       \caption {\small Received illuminance at the location of PD versus perceived brightness percentage.}
                   \vspace{0in}
  \end{figure}

\begin{figure}\label{5}
       \centering
       \includegraphics[width=3.2in]{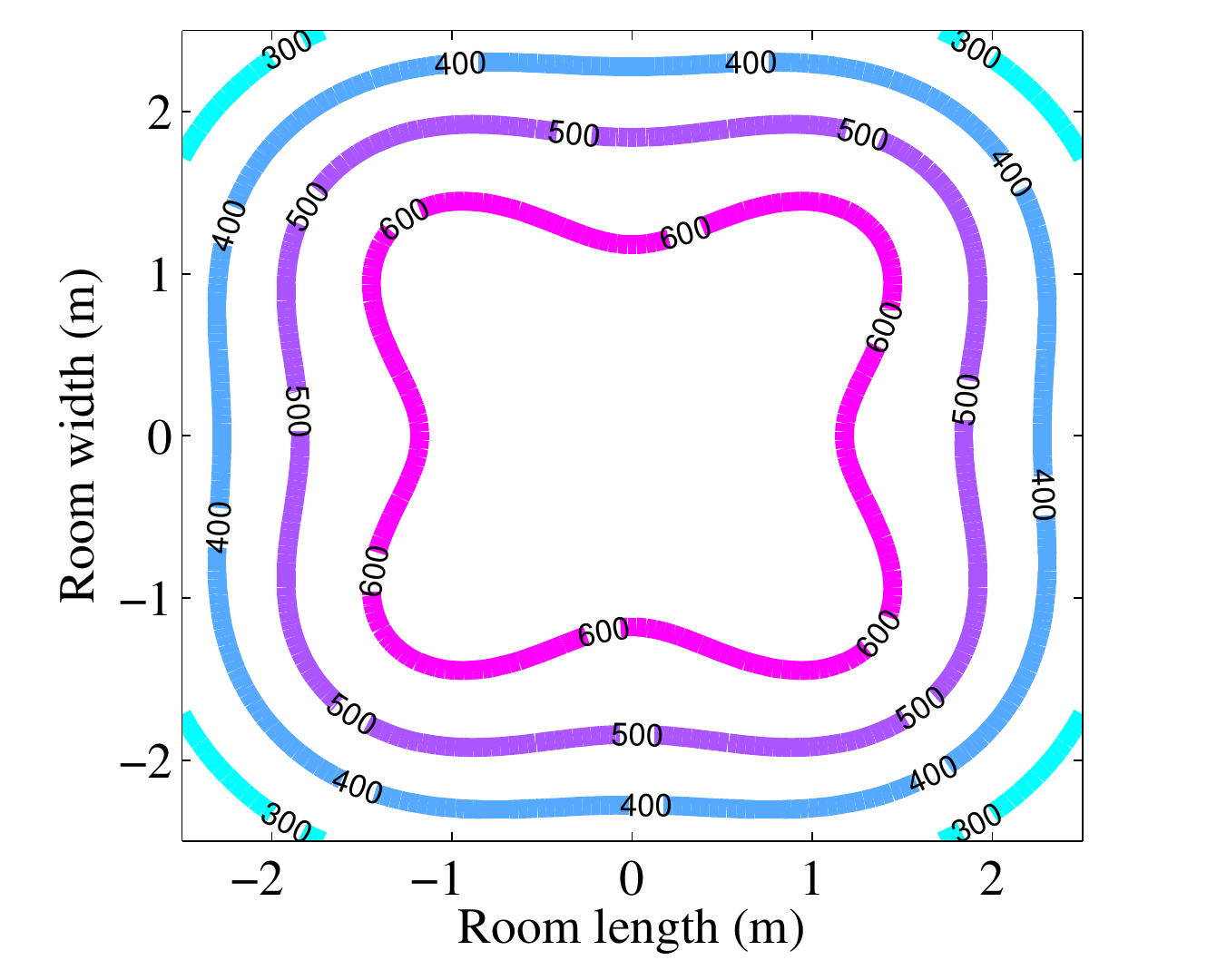}
       \caption {\small Distribution of brightness within the room at $80 \%$ of perceived brightness.}
                   \vspace{-0.1in}
  \end{figure}
  
\subsection{Maximum Achievable Bit Rate}
The bit rate in VLC systems is affected by multipath reflections. An established criterion for defining its upper bound is root mean square (RMS) of delay spread which is given by\cite{ghassemlooy2012optical};
\begin{equation} \label{eq10} 
D=\sqrt{\frac{\int \left(t-\mu \right)^{2} h^2\left(t\right)dt }{\int h^{2} \left(t\right)dt } },
\end{equation}
where $\mu={\int th^{2} (t)dt }/{\int h^{2} (t)dt }$ is the mean delay resulted by NLOS paths and $h(t)$ is the dispersive channel impulse response, described in  \eqref{eq7}. Therefore, to guarantee an ISI-free transmission, the maximum bit rate should be less than $1/(10D)$.
 The delay spread parameter depends on room dimensions, location of transceivers, and also the FOV of PD.
Fig. 6 depicts the distribution of RMS delay spread within the room. This parameter at the position of PD in the hypothetical room is obtained as $1.28$ ns. Hence, the bit rate would be upper bounded by $78$ Mbps. 
On the other hand, the maximum bit rate theoretically achievable for different modulation schemes, namely VOOK, MPPM, and OPPM, versus $44$ to $90$ percent of perceived brightness is illustrated in Fig. 7.
 The OPPM modulation scheme not only gives the highest bit rate (e.g., up to $50$ Mbps compared to MPPM and VOOK \cite{lee2011modulations} which give $20$ Mbps and $7$ Mbps, respectively, at $80 \%$ of perceived brightness), but also referring \eqref{eq5} the peak value of OPPM bit rate, which is $50$ Mbps, is less than the upper bound achieved due to channel constraint, namely $78$ Mbps.
 Note that the modulation bandwidth of LEDs is supposed to be about $20$ MHz, because although the bandwidth offered by off-the-shelf phosphor-based LEDs is about $2$ MHz, it could be improved to $20$ MHz by installing a blue filter on PD in order to remove the yellowish component of the received light \cite{ghassemlooy2012optical}. 
 
\begin{figure}\label{6}
       \centering
       \includegraphics[width=3.6in]{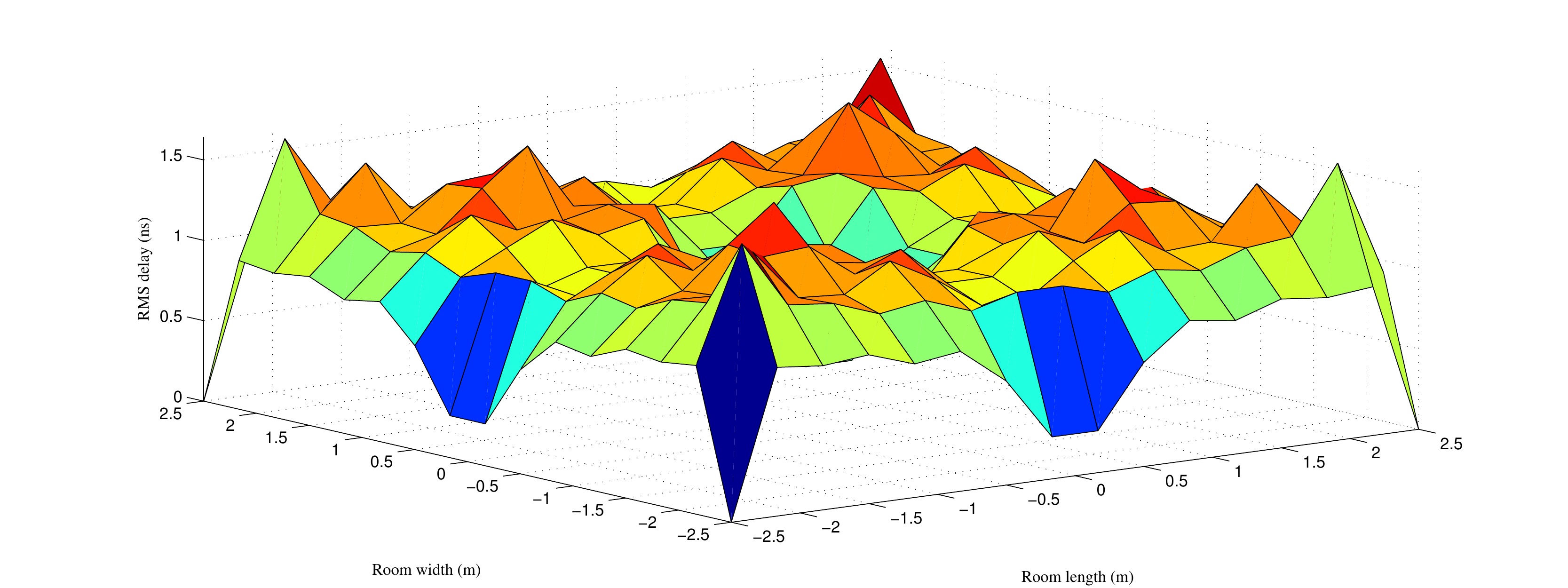}
       \caption {\small Distribution of RMS delay spread within the given room.}
                   \vspace{0in}
  \end{figure}

\begin{figure}\label{7}
       \centering
       \includegraphics[width=3.4in]{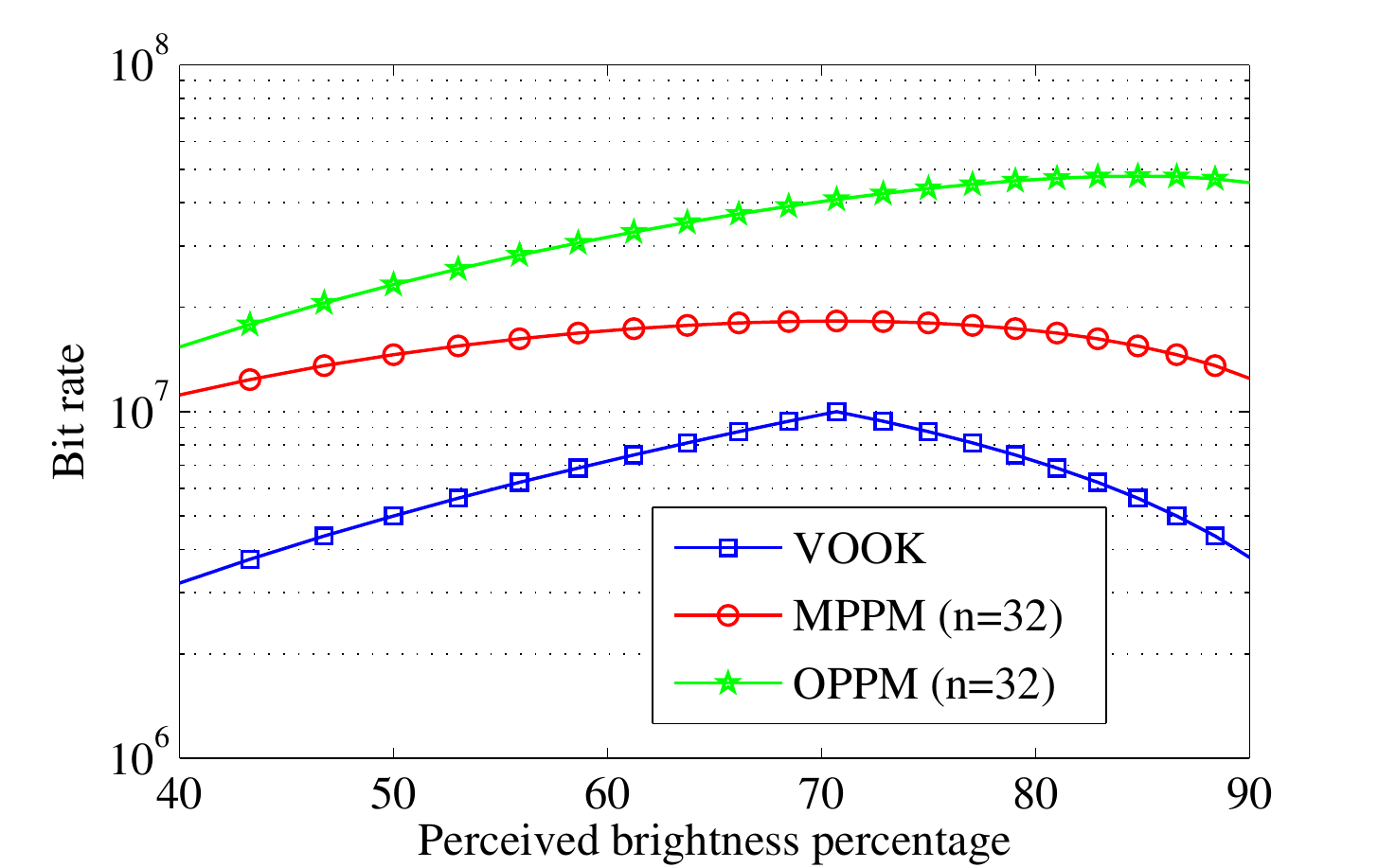}
       \caption {\small Maximum bit rate theoretically achievable in terms of allowable dimming interval \cite{gancarz2015overlapping}.}
                   \vspace{-0.1in}
  \end{figure}
\subsection{Trellis Coded OPPM}	
It is widely known that TCM improves system performance without increasing the required bandwidth. Since OPPM symbols have relatively low duty cycle and equal energy, if the duty cycle $\Delta$ stands fix, then doubling the number of symbols $L$ leads the bandwidth to remain constant \cite{georghiades1989some}. For this reason, the number of chips of the eventuated TCM $2L$-OPPM can be calculated as $n_{c} =\frac{2L-1}{1-\Delta }$ and the equivalent weight would be ${w}_{c}=\Delta {n}_{c}$. However, by applying TCM to OPPM the minimum hamming distance between symbols decreases which causes to a growth in the probability of error. Fortunately, the arisen coding gain via set partitioning retrieves the performance degradation of reduced minimum hamming distance. Fig. 8 outlines the set partitioning of the OPPM with $n=16$. The bandwidth requirement of the TCM $2L$-OPPM is the same as uncoded OPPM and can be calculated using \eqref{eq4.5}, yet the required average power could be lessened as is described in the following.

Considering the minimum hamming distance as ${d}=\frac{P}{w} \sqrt{dnT}=2$ for uncoded OPPM \cite{park1995modulation}, the BER would be;
\begin{equation} \label{EQ13} 
BER=Q(\frac{{P}\sqrt{dnT} }{2w\sqrt{N_{0} } } ). 
\end{equation} 
By solving the last equation for $P$, the average required optical power of the uncoded OPPM can be extracted as;
\begin{equation} \label{EQ14} 
P_{uc}^{(\rm OPPM)}=2w\sqrt{\frac{N_{0} }{2nT} } Q^{-1} (BER).
\end{equation} 
Nevertheless, for TCM $2L$-OPPM we have;
\begin{equation} \label{EQ15} 
P_{\rm TCM}^{(2L-{\rm OPPM})} =2\Delta w\sqrt{\frac{N_0}{d_{c} \frac{2L-1}{1-\Delta } T} } Q^{-1} (BER),
\end{equation} 
where ${d}_{c}$ is the minimum hamming distance between TCM $2L$-OPPM symbols. Therefore, the consequent coding gain, in terms of dB, is given by;
\begin{align}\label{EQ16} 
\frac{P_{uc}^{(\rm OPPM)}}{P_{\rm TCM}^{(2L\!-\!{\rm OPPM})}}\!\!=\!\!10\!\log_{10}\!\!\left(\!\frac{L\!-\!1}{2L\!-\!1}\sqrt{\!\frac{d_c}{2}\frac{2L\!-\!1}{L\!-\!1}}\right)\!\!\approx\!\! 10\!\log_{10}\!\!\left(\!\!\sqrt{\frac{d_c}{4}}\right)\!.
\end{align}
For the sake of achieving the coding gain, ${d}_{c}$ needs to be greater than four, e.g., examine $9$-OPPM with $w=8$. A coding gain equal to $1.2$ dB is attainable by choosing a $4$-state set partitioning of the TCM $18$-OPPM with $w_c=17$ along with the same bandwidth requirement. Altogether, the superiority of trellis coded OPPM over uncoded OPPM is depicted in Fig. 9, in which the power requirement to achieve a BER of ${10}^{-6}$ through the given room is plotted in terms of the qualified dimming interval. As mentioned in \eqref{EQ16}, although the required power for uncoded and TCM $2L$-OPPM with ${d}_{c}=4$ are the same, increasing ${d}_{c}$ to $8$ and $16$ leads to a reduction of $2.55$ dB and $3$ dB in power requirement, respectively with the same bit rate.
\begin{figure}[t]\label{8}
       \centering
       \includegraphics[width=3.6in]{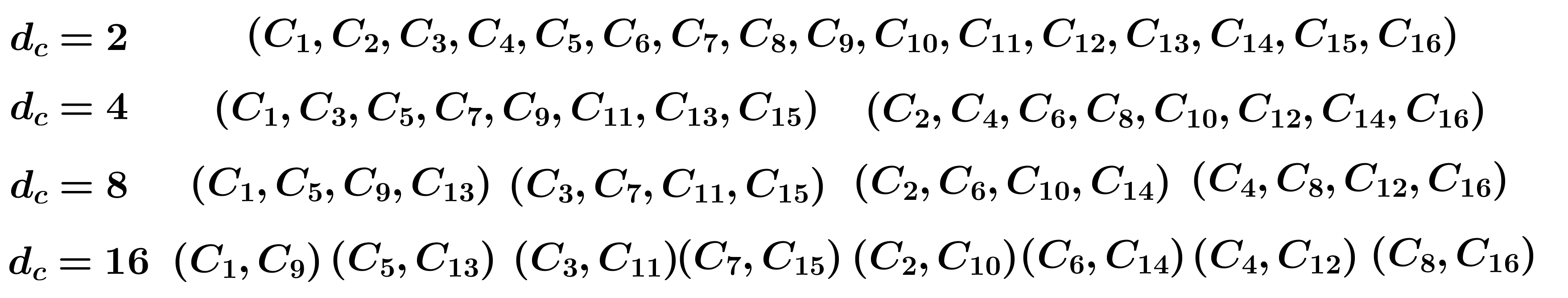}
       \caption {\small 1, 2, 4, and 8-state set partitioning.}
                   \vspace{-0.1in}
  \end{figure}
\begin{figure}\label{9}
       \centering
       \includegraphics[width=3.4in]{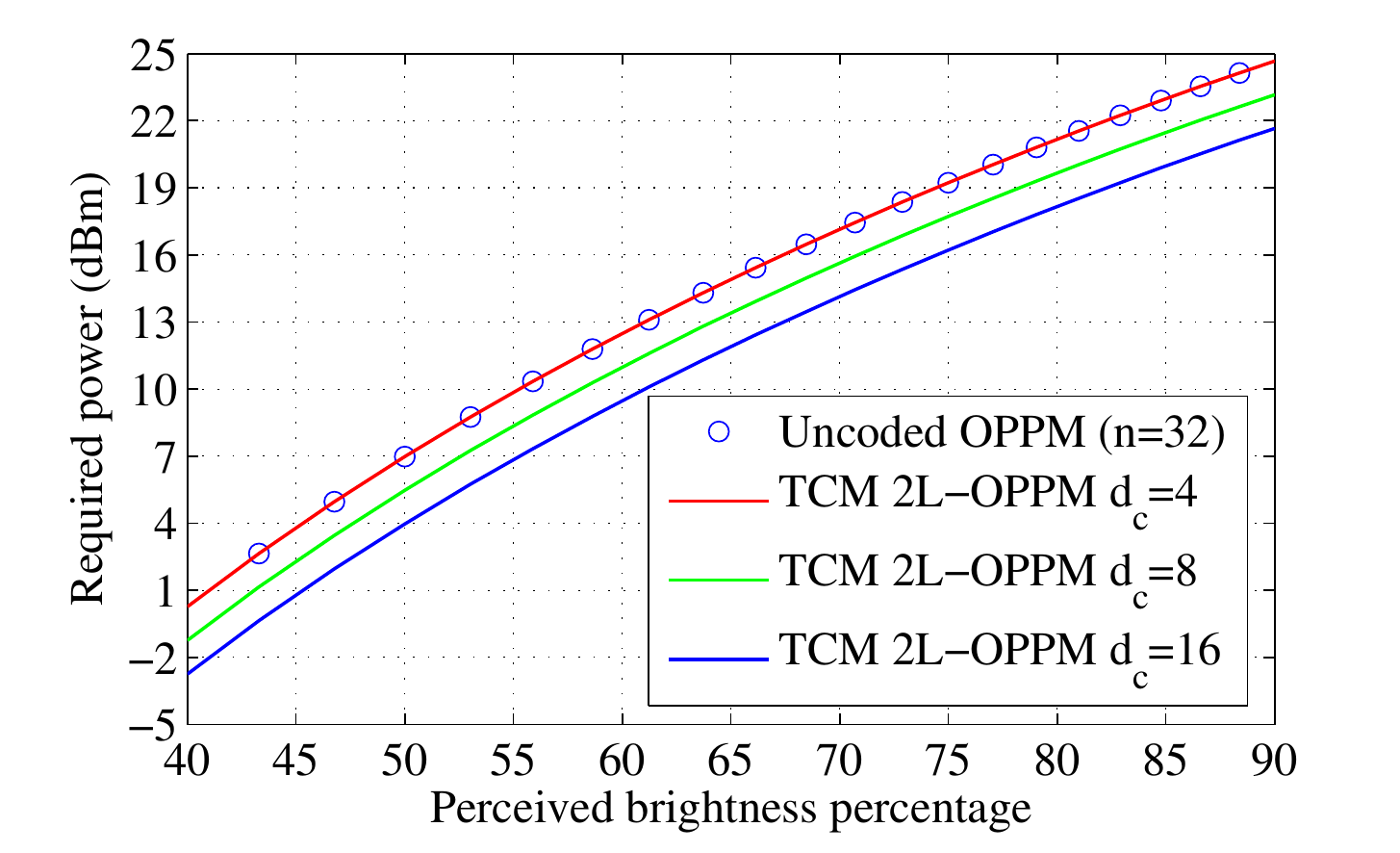}
       \caption {\small Average optical power requirement of OPPM (${BER=10}^{-6}$) in terms of allowable dimming interval at the location of PD within the hypothetical room.}
                   \vspace{-0.1in}
  \end{figure}

\subsection{Minimum Required Code Length}
As the third part of system design considered in this paper, we need to analyze the system from the BER viewpoint. Referring \eqref{EQ13} and assuming $d=2$ for uncoded OPPM and also $T=\log_2^L/R_b$, one can write;
\begin{equation} \label{EQ17} 
BER=Q(\frac{P}{w}\sqrt{ \frac{n\log_2^L} {2R_b{N_{0} } }} ). 
\end{equation} 
In this equation, $n$ is the code length which needs to be chosen in order to achieve an acceptable bit error probability. To do so, the maximum BER of $3\times 10^{-3}$ as a threshold for forward error correction (FEC) is supposed. The parameter $N_0$ in \eqref{EQ17} contains the contributions of shot noise, thermal noise, and multipath reflections. Therefore, we simulate the system shown in Fig. 1 in two steps.
 
\noindent{\textit{Step I:}} A transmitter containing OPPM encoder sends modulated symbols through the channel in which the noise is considered to be AWGN. At the receiver side the temporal data streams after passing through a filter which is matched with the transmitted pulse shape, $p(t)$, and being sampled with rate $1/T$, are detected using a hard decision block. Finally, the bit error probability is calculated using MC simulation, in terms of different SNRs. This procedure is repeated for $4$ dimming levels, namely $35$, $50$, $75$, and $86\%$ which are illustrated in Fig. 10, respectively. Furthermore, in each dimming level, such simulation is done with $5 $ different code lengths.

\noindent{\textit{Step II:}} In this step, to consider channel and receiver constraints, including the effects of multipath reflectors, signal dependent shot noise, and thermal noise \cite{komine2004fundamental}, the ratio of received optical power to sum of noise contributions is simulated within the hypothetical room to determine the channel SNR according to Eq. \eqref{eq6}.
 Fig. 11 depicts the SNR in terms of a range of dimming levels. As a consequence, to choose an appropriate code length for a specific dimming level, one can extract the amount of channel SNR in order to determine the probability of error from the corresponding curve (Fig. 10). For instance, for $50 \%$ of dimming, the channel SNR using Fig. 11 would be $-0.5$ dB. Then, regarding Fig. 10(b) the code lengths $32$, $64$, and $128$ have BER values less than the predetermined BER threshold, i.e., $3\times 10^{-3}$. Thus, the minimum code length of $32$ can be used in this dimming level. Table II summarizes  minimum code length appropriate for different levels of dimming.

\begin{table}[t]
\centering
\caption{\small Minimum acceptable code length for various perceived brightness percentages.}
\label{II}
\begin{tabular}{|c|c|c|}
\hline
{\bf Dimming level (\%)}& {\bf Channel SNR (dB)} &{\bf Minimum acceptable $n$} \\ \hline
$35$                 			&			$-6.2$				& $128$ \\ \hline
$50$                 			&			$-0.5$				& $32$ \\ \hline
$75$                 			& 			$6.5$				&$8$ \\ \hline
$86$                 			&			$8.5$				& $8$ \\ \hline
\end{tabular}
\end{table}
\begin{figure*}
    \centering
    \begin{subfigure}[b]{0.4\textwidth}
        \centering
        \includegraphics[width=\textwidth]{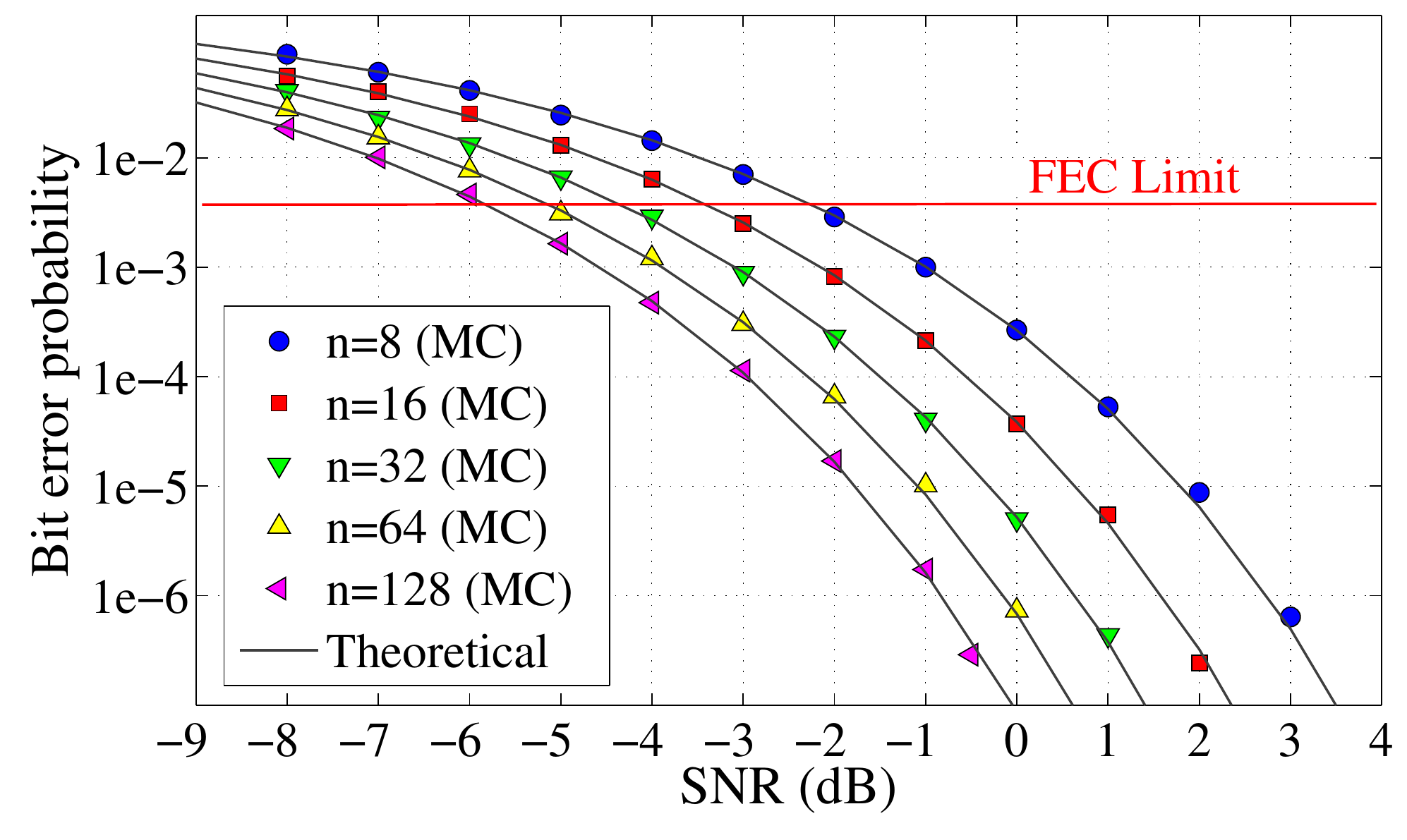}
        \caption{\small $35 \%$}
        \label{a}
    \end{subfigure}
    \begin{subfigure}[b]{0.4\textwidth}
        \centering
        \includegraphics[width=\textwidth]{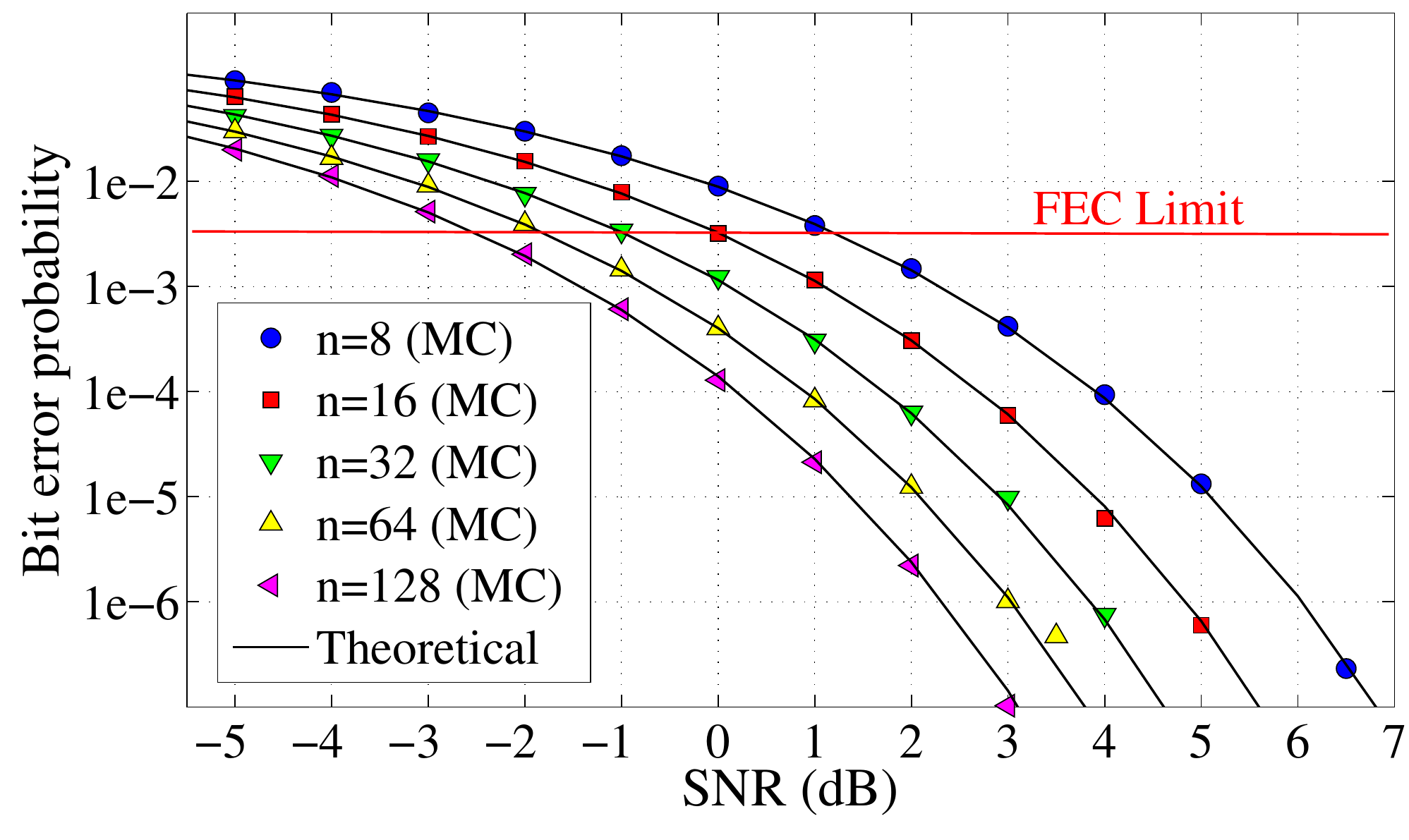}
        \caption{\small $50 \%$}
        \label{b}
    \end{subfigure}
    \begin{subfigure}[b]{0.4\textwidth}
        \centering
        \includegraphics[width=\textwidth]{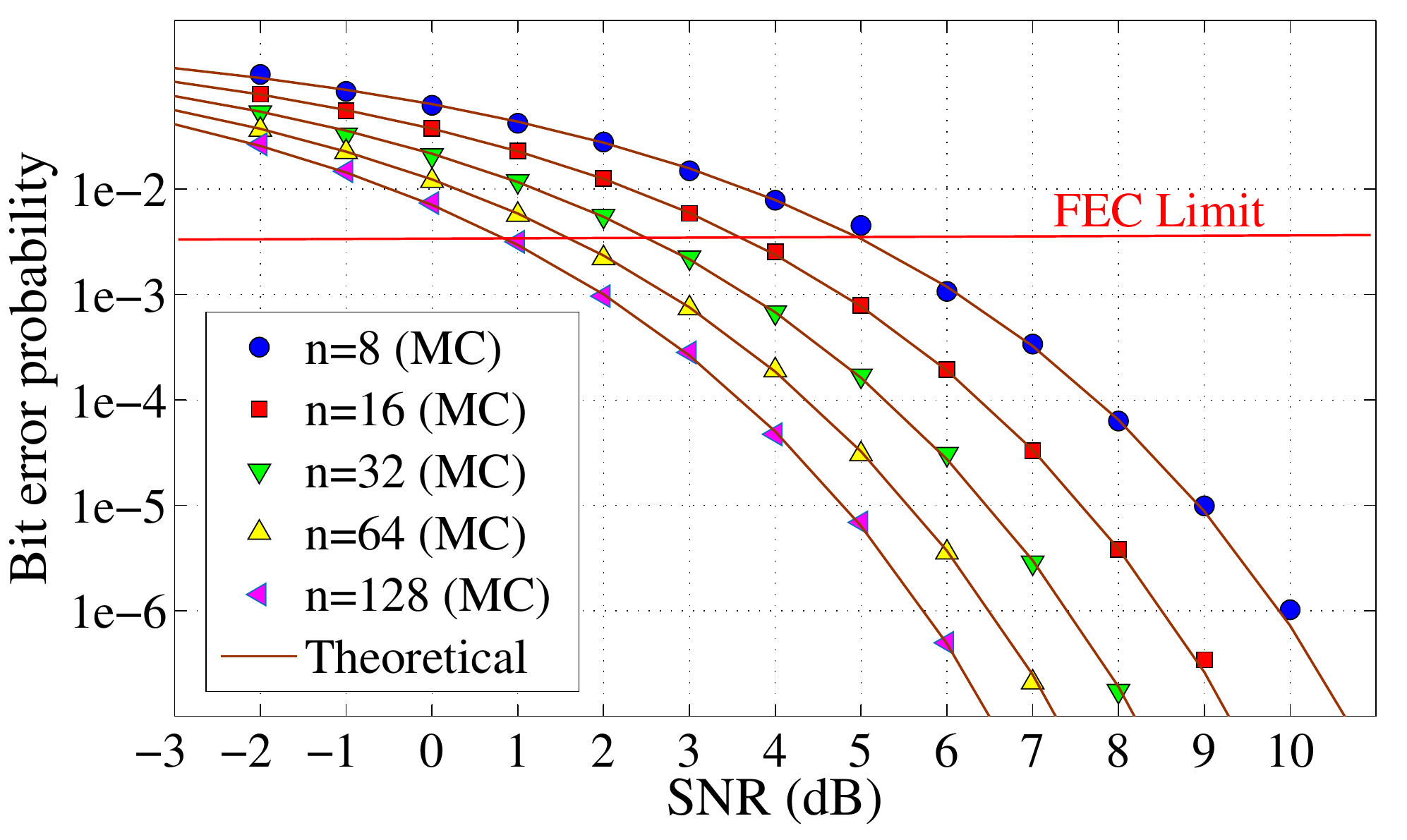}
        \caption{\small $75 \%$}
        \label{c}
    \end{subfigure}
        \begin{subfigure}[b]{0.4\textwidth}
            \centering
            \includegraphics[width=\textwidth]{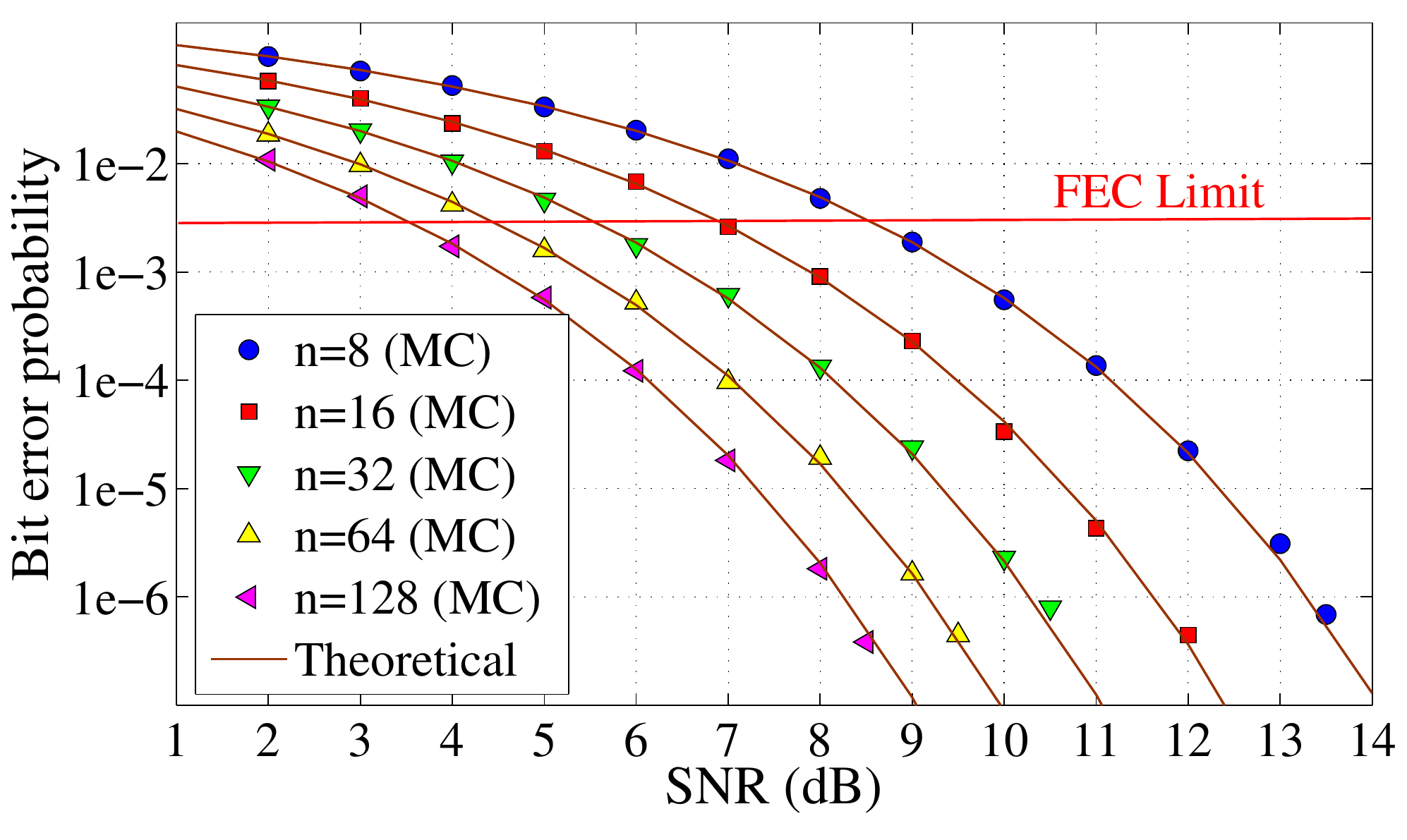}
            \caption{\small $86 \%$}
            \label{d}
        \end{subfigure}
    
    \caption{\small Bit error probability of OPPM versus SNR for various perceived brightness percentages.}
    \label{11}
    \vspace{-0.1in}
\end{figure*}
  \begin{figure}\label{14}
	             \centering
	             \includegraphics[width=3.4in]{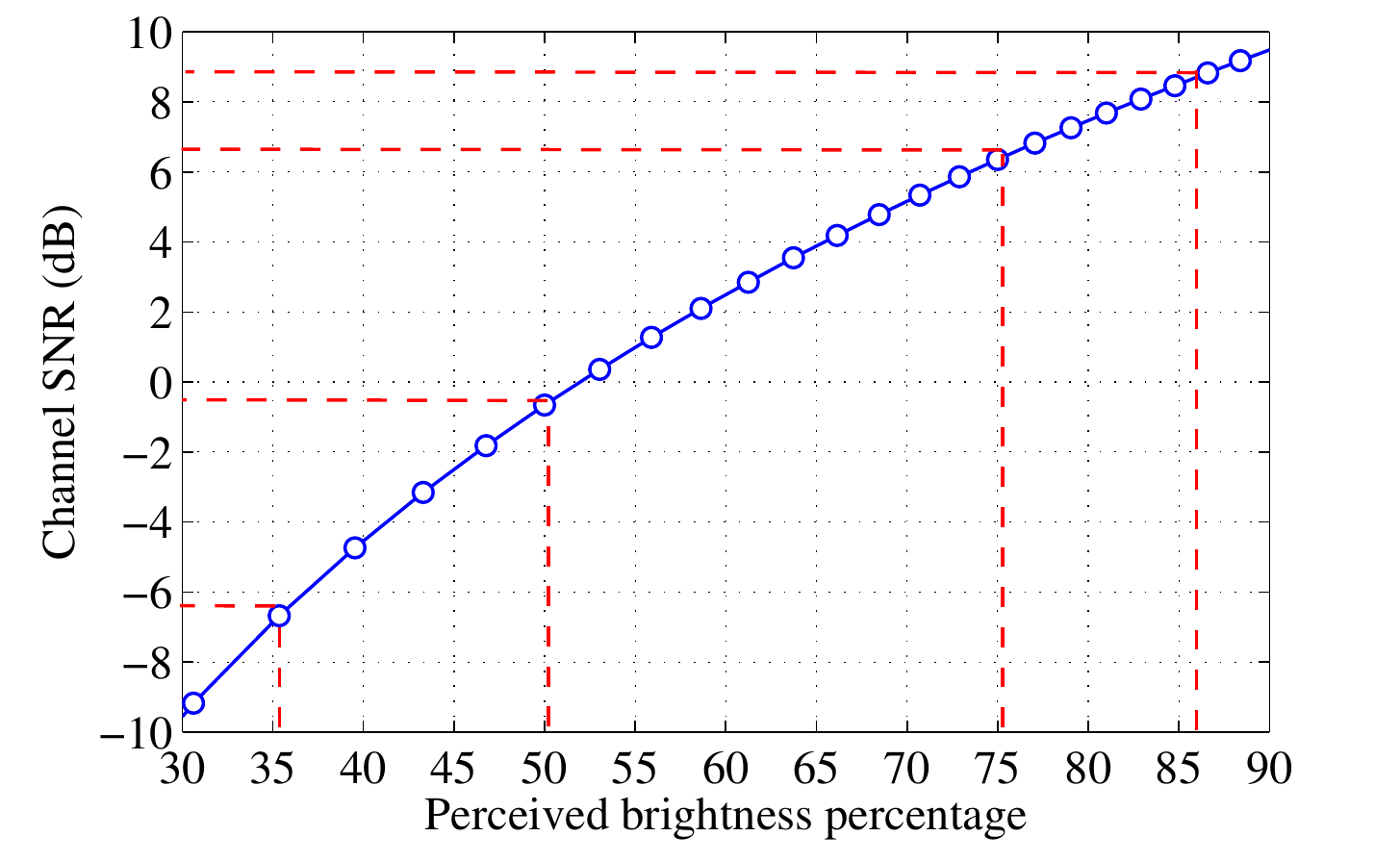}
	             \caption {\small Simulation of channel SNR within the hypothetical room at the location of PD for different levels of perceived brightness percentages.}
	                         \vspace{-0.1in}
	        \end{figure}	
\section{Conclusion}	
 In this research, we investigated the design of a VLC system which supports dimming. We exploited OPPM due to its less power requirement in comparison to MPPM. The dimming control was addressed by changing the code weights while remaining the code lengths unchanged. At first, the dimming interval of $44$ to $90$ percent was determined according to illumination standards in a hypothetical office room. Then, an upper bound of $78$ Mbps for ISI-free transmission was obtained via simulation of dispersive channel. Moreover, TCM was suggested to be applied to OPPM in order to take advantage of about $3$ dB coding gain. Finally, the minimum code length that achieves a minimum BER for different dimming levels was obtained via MC simulation.
	
%


\end{document}